# Apply Sorting Algorithms to FAST Problem

Sadra Mohammadshirazi
Dept. of Mathematics and Computer Science
Amirkabir University of Technology
Tehran, Iran
sadra.ms69@aut.ac.ir
sadra.ms69@gmail.com

Alireza Bagheri
Dept. of Computer Engineering
Amirkabir University of Technology
Tehran, Iran
ar_bagheri@aut.ac.ir

## ABSTRACT
FAST problem is finding minimum feedback arc set problem in tournaments. In this paper we present some algorithms that are similar to sorting algorithms for FAST problem and we analyze them. We present Pseudo_InsertionSort algorithm for FAST problem and we show that average number of all backward edges in output of that is equal to $\frac{n^2 - 5n + 8}{4} - 2^{1-n}$. We introduce Pseudo_MergeSort algorithm and we find the probability of being backward for an edge. Finally we introduce other algorithms for this problem.

## 1. INTRODUCTION
Suppose a chess tournament that each player should exactly play chess with other players exactly one time. So we have $k \times (k-1)/2$ games if there are *k* players. After all games finish if we want to have a ranking that is best we should order players in which number of pair players for example player A and player B which A defeat B but A is in worse rank should be minimized. This problem is related to FAST problem. If we create a graph which each vertices of the graph should assign to each player and the directed edge between each pair nodes is from winner to looser then solving FAST problem is equal to finding optimal ranking.

FAST problem has other applications. Suppose there are *n* judges and each judge make a ranking on *k* candidates. So we have *n* permutations on *k* candidates and we want to find one optimal ranking of these *k* candidates which sum of distances between the optimal ranking and n input ranking is minimal. This problem is known as rank aggregation problem.

Actually this paper applies sorting algorithms to FAST problem. FAST problem is minimum feedback arc set in tournaments which tries to find order of vertices in an input tournament that minimizes backward edges. For example consider following tournament:

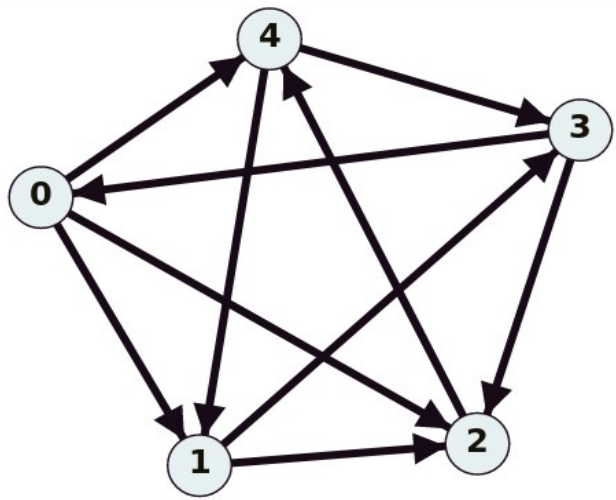

For above tournament we consider an order of vertices for example "0,1,2,4,3". In the order the edge from 3 to 0 is a backward edge. The edge from 4 to 1 is also a backward edge. The edge from 3 to 2 is a backward edge. So the order "0,1,2,4,3" has 3 backward edges and cost of the order is 3. Sorting algorithms define on numbers for example if we have an input array of numbers like "3,6,7,1,9,4" and we apply a sorting algorithm on the array then we will have "1,3,4,6,7,9" as output. In this problem we have a tournament instead of numbers and we must define largeness and smallness in a tournament if we want to use something like sorting algorithms here. Consider following edge:

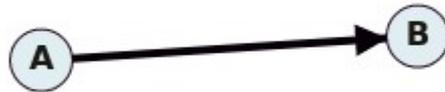

In the above edge because the edge is from A to B we consider B is larger than A and we try to produce an order as output that B comes after A, something like …A…B….
In this paper we introduce some algorithms that is not exactly the sorting algorithm which is known but they are similar to them. For example in Pseudo_InsertionSort algorithm first we consider 2 vertices like A and B:

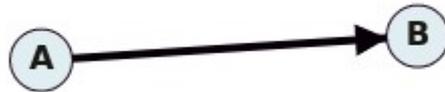

Because the edge is from A to B then we put A in first place in output order and B in second place in output order. It is the same as choosing first two numbers in insertion sort algorithm. Then we consider next vertex for example C. Now we compare C to B, if the edge is from B to C then we put C in third place in output order.

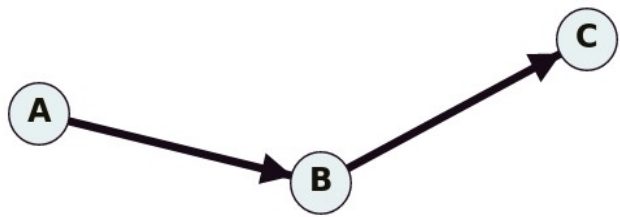

If the edge is from C to B then we compare C to A and if the edge is from A to C in this situation then we put C in second place in output order and B in third place in output order.

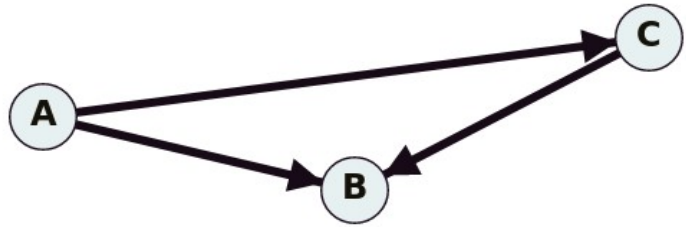

If the edge is from C to A then we put C in first place and A in second place and B in third place.

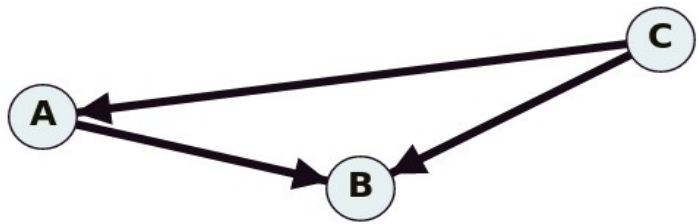

It is the same as insertion sort when we choose third number and we want to find correct place of it in output order. We continue this process until we put all vertices in output order.

Another algorithm for FAST problem is Pseudo_QuickSort Algorithm which first choose a random vertex A and put all vertices which their edges are from them to A previous A. We put other vertices which their edges are from A to them after A. Then we do same process recursively for the group vertices before A and the group vertices after A separately and we return final order as output.

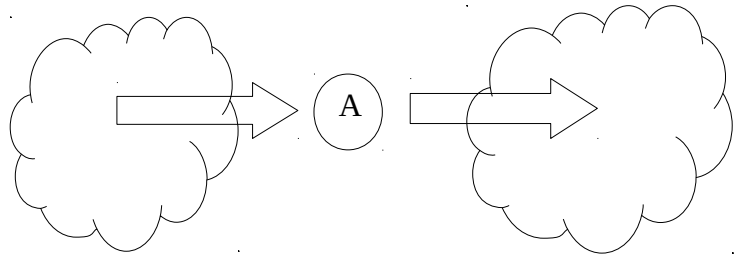

Ailon, Charikar and Newman [*] proved that Pseudo_QuickSort algorithm is a randomized expected 3-approximation algorithm. Minimum feedback arc set problem can be defined on general directed graphs (not just tournaments) and in [*,*] presented an approximation algorithm for that. It was proved that approximation hardness of minimum feedback arc set problem is 1.36 [*,*]. Mathieu and Schudy presented a polynomial time approximation scheme (PTAS) for the minimum feedback arc set problem on tournaments [*].

## 1.1 Organization

In section 2 we introduce Pseudo_InsertionSort Algorithm and we show the average number of all backward edges in output of that. In section 3 we introduce Pseudo_MergeSort Algorithm and show the probability of an edge being backward. In section 4 we introduce some other pseudo sorting algorithms.

## 1.2 Preliminaries and definition

D-graph: A d-graph is a set of vertices V and a set of edges E which is ordered pair of $v \in V$ that connect the vertex in first place of the pair order to the vertex in second place of the pair order.

Tournament: A tournament is a d-graph which each pair of $v \in V$ has exactly one edge.

Backward edge: In an order of V of a tournament if $x \in V$ is before $y \in V$ and the direction of edge that connect y to x is from y to x then the edge is called backward edge.

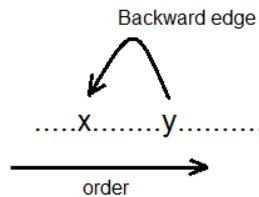

FAST problem: FAST problem is minimum feedback arc set in tournament that tries to find an order of V that minimize backward edges.

## 2. PSEUDO_INSERTIONSORT ALGORITHM

This algorithm is like insertion sort algorithm and it is randomized version of that. First we have a list of all input vertices named V. We choose a random vertex v(1) and then another random vertex v(2) from the list we compare them and if the edge between them is from v(1) to v(2) then we do nothing but if the edge is from v(2) to v(1) we change the place of v(1) and v(2) and put v(2) in first and v(1) in second place. Then we choose v(3) randomly and do the same process like insertion sort until all vertices finished.

```
Pseudo_InsertionSort (G= (V, E))
    Let V is input list of vertices
    Randomly permute V
    For (i=2 to n)
        j=i
        While (V[j] → V [j-1])
            Swap (V[j], V [j-1])
            j=j-1
Return V
```

Lemma 2: By swapping each adjacent element in each order of vertices we can make each order of vertices that exist.

Proof: Assume we want to swap a(i) and a(j) in a(1), a(2), a(3), …, a(i), …, a(j), …, a(n) without changing in the place of other elements. First we change the place of a(i) with next vertices and we do it until a(i) reach the a(j) and be placed in a(j) like a(1), …, a(i-1), a(i+1), a(i+2), …, a(j-1), a(j), a(i), a(j+1), …,a(n). Now we swap a(j) with previous elements until it placed between a(i-1) and a(i+1). Now we prove that by swapping each adjacent element in permutation P we can swap each arbitrary pair of elements without changing place of other elements. Assume two following permutations: a(1), a(2), …, a(n) and b(1), b(2), …, b(n). We want to make one of them by changing the place of adjacent elements in the other one. First we find b(1) in first permutation and we swap it with a(1), now we reach b(1), a(2), a(3), …, a(1), ….,a(n). Next we do this process with b(2) and after that b(3) until b(n) now we reach b(1), b(2), …, b(n).

Lemma 3: The output of Pseudo_InsertionSort is in local minimum.

Proof: Suppose that it is not in local minimum, so by swapping adjacent elements we can reach better order, for example suppose that the output of algorithm is a(1), a(2), …, a(i), a(i+1), …, a(n) and suppose that if we swap a(i) and a(i+1) we can reach better score, so we have a(i+1)→a(i) and that is contradiction due to the algorithm because assume that in the $i^{th}$ stage of algorithm the element b(i) is in $i^{th}$ place and element b(i+1) is in $(i+1)^{th}$ place, in this stage b(i) and b(i+1) compared and if we have b(i+1)→b(i) then they would swap and if we use induction assume that in stage k of the algorithm which k≥(i+1) we have c(i) in i[th] place and we have c(i+1) in (i+1)[th] place and we have c(i)➔c(i+1) so we prove that for stage k+1. In stage k+1 assume that we have d(k+2) in (k+2)[th] place, we have d(k+1) in (k+1)[th] place, d(i) in i[th] place and d(i+1) in (i+1)[th] place.

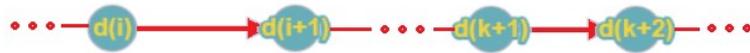

If the order like above, the direction of edge between d(k+1) and d(k+2) is d(k+1)➔d(k+2) then we do not have change in the order and so we have d(i)➔d(i+1) so in this situation the proof is complete.

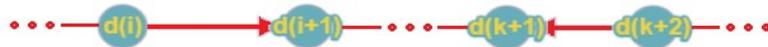

If the order like above, the direction of edge between d(k+1) and d(k+2) is d(k+2)➔d(k+1) then we swap d(k+1) with d(k+2), d(k+2) move to left and if d(k+2) does not pass d(i+1) then we have d(i)➔d(i+1) so in this situation the proof is complete, and if d(k+2) passes d(i+1) and also passes d(i) then we have d(i)➔d(i+1), if d(k+2) passes d(i+1) but does not pass d(i) like below it is d(i)➔d(k+2) so the proof is complete.

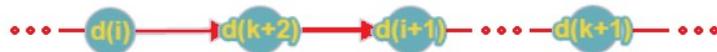

Theorem 1: The average number of all backward edges in output of Pseudo_InsertionSort algorithm is equal to $\frac{n^2-5n+8}{4}-2^{1-n}$

Proof: Assume that B(k) is the number of backward edges from the vertex that we put in the order at stage k.

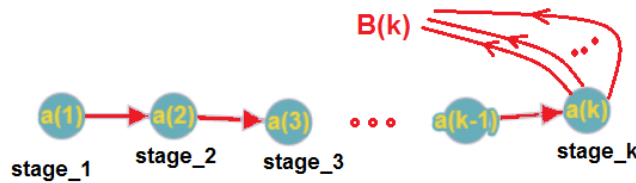

It is clear that B(1)=0 and B(2)=0 and we have $B(k)=\frac{B(k-1)}{2}+\frac{k-2}{4}$, because if we have a(k-1) ➔ a(k), (the direction of edge is from a(k-1) to a(k)) so we have $\frac{k-2}{2}$ backward edges because there are k-2 vertices which each of them with probability of 0.5 have edge toward it from a(k), and the probability of that is 0.5 and if we have a(k) ➔ a(k-1) then we have B(k-1) backward edges and the probability of that is 0.5. If we solve the relationship of B(k) then we find that $B(k)=\frac{k}{2}-\frac{3}{2}+2^{1-k}$.

The average number of all backward edges in output of the algorithm is equal to $\sum_{k=3}^{n} B(k)=\frac{n^2-5n+8}{4}-2^{1-n}$.

## 3. PSEUDO_MERGESORT ALGORITHM

This algorithm is like merge sort algorithm. We pick $\lfloor \frac{n}{2} \rfloor$ vertices randomly from V and run the algorithm on it then we pick other $\lceil \frac{n}{2} \rceil$ vertices from V and run the algorithm on it. Next we merge the two groups.

```
Pseudo_MergeSort (G= (V, E))
    Assume V=V[1],V[2],…,V[n]
    Select ⌊n/2⌋ vertices from V randomly and put it in VF
    Put other vertices in VS
        OF = Pseudo_MergeSort (G= (VF, EF)) where EF ⊂ E such that endpoints of e ∈
        EF are in VF
        OS = Pseudo_MergeSort (G= (VS, ES)) where ES ⊂ E such that endpoints of e ∈
        ES are in VS
        Let kf=1, ks=1, list O=null
        For (i=1 to n)
            If(OF[kf]→OS[ks])
                insert OF[kf] to O
                kf++
            If(OF[kf]←OS[ks])
                insert OS[ks] to O
                ks++
            if(kf==⌊n/2⌋ or ks==n) break
        Put other vertices to O with its order
Return O
```

Assume that we have two groups of vertices which each one has n vertices. We define H(i,j) that show the probability of comparison of $i^{th}$ vertex in first group to $j^{th}$ vertex in second group.

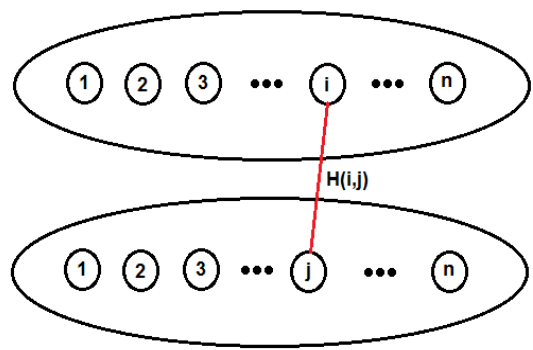

Lemm 4: H(i,j) is equal to H(j,i)
Proof: because we have symmetric in this situation those are equal.

Theorem 2: $H(i,j) = \binom{i+j-2}{i-1} \times \frac{1}{2^{i+j-2}}$

Proof: we have $H(i,j) = H(i,j-1) \times \frac{1}{2} + H(i-1,j) \times \frac{1}{2}$ by its definition, because $i^{th}$ vertex in first group compared with $(j-1)^{th}$ vertex in second group with probability of H(i,j-1) and the direction of edge is from $(j-1)^{th}$ vertex in second group to $i^{th}$ vertex in first group with probability of 0.5.

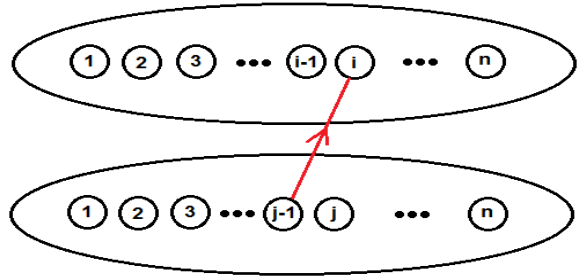

We have $H(1,n)=H(n,1)=(\frac{1}{2})^{n-1}$ and by solving the equation we have $H(i,j)=\frac{i(i+1)\ldots(i+j-2)}{(j-1)!\times 2^{i+j-2}}$

and finally we have $H(i,j)=\binom{i+j-2}{i-1}\times\frac{1}{2^{i+j-2}}$

**Theorem 3:** Assume ij is the edge that connect $i^{th}$ vertex in first group to $j^{th}$ vertex in second group. Let $P(y_{ij}=1)$ is the probability of the edge ij being backward, we have

$$P(y_{i,j}=1)=(\frac{1}{2})^2\times\left[\sum_{k=1}^{j-1}\binom{k+i-2}{i-1}\times\frac{1}{2^{k+i-2}}+\sum_{k=1}^{i-1}\binom{k+j-2}{j-1}\times\frac{1}{2^{k+j-2}}\right]$$

Proof:

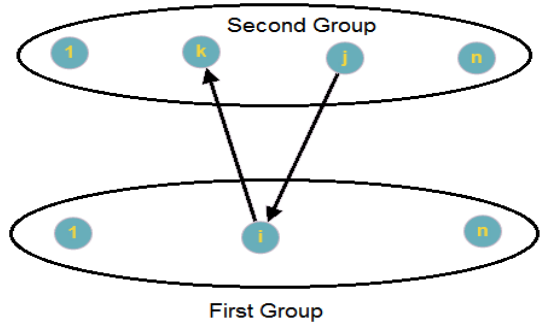

In the above graph with probability 0.5 we have j→i and with probability 0.5 we have i→k and with probability $H(k,i)$ vertices k and i are compared to each other. So with probability $0.5\times 0.5\times H(k,i)$ the edge ji is backward. So we have:

$$P(y_{i,j}=1)=\frac{1}{2}\times\left[H(1,i)\times\frac{1}{2}+H(2,i)\times\frac{1}{2}+\ldots+H(j-1,i)\times\frac{1}{2}\right]+\frac{1}{2}\times\left[H(1,j)\times\frac{1}{2}+H(2,j)\times\frac{1}{2}+\ldots+H(i-1,j)\times\frac{1}{2}\right]$$

$$P(y_{i,j}=1)=(\frac{1}{2})^2\times\left[\sum_{k=1}^{j-1}H(k,i)+\sum_{k=1}^{i-1}H(k,j)\right]$$

$$P(y_{i,j}=1)=(\frac{1}{2})^2\times\left[\sum_{k=1}^{j-1}\binom{k+i-2}{i-1}\times\frac{1}{2^{k+i-2}}+\sum_{k=1}^{i-1}\binom{k+j-2}{j-1}\times\frac{1}{2^{k+j-2}}\right]$$

## 4. OTHER PSEUDO_SORTING ALGORITHMS

**Lemma 1:** Suppose the order of vertices a(1), a(2), …, a(n) is optimal order, for $1\leq i\leq n-1$ the direction of edge is from a(i) to a(i+1).

Proof: Suppose there are two vertices a(k) and a(k+1) which the direction of edge is from a(k+1) to a(k) so we change the place of a(k) and a(k+1) and we make the order a(1), a(2), …, a(k+1), a(k), …., a(n). The order has one backward edge less than optimal order so it is contradiction.

## 4.1. Pseudo_ SelectionSort Algorithm

This algorithm is like selection sort algorithm. In this algorithm we find the vertex that has maximum output degree, if there is more than one vertex which has maximum degree we select one of them randomly. Then we put the vertex at first of the order. We continue this process until all vertices selected.

```
Pseudo_ SelectionSort (G= (V,E))
    Set V_output=0
        While (G is not empty)
            Select a vertex v in V which has maximum output degree
            Add v to V_output
            Remove v from G
    Return V_output
```

## 4.2. Pseudo_BubbleSort Algorithm

This algorithm is like bubble sort algorithm. First we have a list of all input vertices in a list named V. For two first vertices V[1] and V[2] in the list we compare them and if the edge between them is from V[1] to V[2] then we do nothing but if the edge is from V[2] to V[1] we swap the content of V[1] and V[2]. Then we compare V[2] and V[3] and do it until all vertices finished. After that we repeat this procedure until no vertices changed.

```
Pseudo_BubbleSort (G= (V, E))
    Let V is input list of vertices
    While (not Swapped)
        For (i=1 to n-1)
        If (V[i+1] → V [i])
            Swap (V[i+1], V [i])
    Return V
```

## 4.3. Pseudo_QuickSort Algorithm

This algorithm is like quick sort algorithm. First we pick a random vertex $v \in V$ then we put all vertices $b \in V$ such that b→v before v and named this set of vertices $V_b$. Next we put all vertices $a \in V$ such that v→a after v and named this set of vertices $V_a$. Now we have $V_b v V_a$ as output and do the same process recursively on $V_b$ and $V_a$. It was proved that this algorithm is a randomized expected 3-approximation algorithm [2].

```
Pseudo_QuickSort (G= (V, E))
    Select a random vertex v in V
    While (not Swapped)
        For (i=1 to n-1)
        If (V[i+1] → V [i])
            Swap (V[i+1], V [i])
    Return V
```

## 5. OPEN PROBLEMS AND FUTURE WORK

Finding approximation factor or expected approximation factor for Pseudo_ SelectionSort Algorithm, Pseudo_InsertionSort Algorithm, Pseudo_BubbleSort Algorithm and Pseudo_MergeSort Algorithm can be future work.
What is the suitable pivot for Pseudo_QuickSort Algorithm?

What is the expected approximation factor of Pseudo_QuickSort Algorithm if we set a vertex $v \in V$ which the value of $|degreeOut(v) - degreeIn(v)|$ is minimal in the group which the algorithm runs on as a pivot?

# 6. ACKNOWLEDGMENTS
I would like to thank Professor Hamid Zarrabi-Zadeh from Sharif University of Technology for a discussion on the problem.